# SLHCat: Mapping Wikipedia Categories and Lists to DBpedia by Leveraging Semantic, Lexical, and Hierarchical Features

Zhaoyi WANG, Zhenyang ZHANG, Jiaxin QIN*, Mizuho IWAIHARA

Graduate School of Information, Production, and Systems, Waseda University
Kitakyushu 808-0135, Japan
`wangzy-joey@akane.waseda.jp zhangzhenyang@fuji.waseda.jp`
`jiaxinqin@fuji.waseda.jp iwaihara@waseda.jp`

**Abstract.**
Wikipedia articles are hierarchically organized through categories and lists, providing one of the most comprehensive and universal taxonomy, but its open creation is causing redundancies and inconsistencies. Assigning DBPedia classes to Wikipedia categories and lists can alleviate the problem, realizing a large knowledge graph which is essential for categorizing digital contents through entity linking and typing. However, the existing approach of CaLiGraph is producing incomplete and non-fine grained mappings. In this paper, we tackle the problem as ontology alignment, where structural information of knowledge graphs and lexical and semantic features of ontology class names are utilized to discover confident mappings, which are in turn utilized for finetuing pretrained language models in a distant supervision fashion. Our method SLHCat consists of two main parts: 1) Automatically generating training data by leveraging knowledge graph structure, semantic similarities, and named entity typing. 2) Finetuning and prompt-tuning of the pre-trained language model BERT are carried out over the training data, to capture semantic and syntactic properties of class names. Our model SLHCat is evaluated over a benchmark dataset constructed by annotating 3000 fine-grained CaLiGraph-DBpedia mapping pairs. SLHCat is outperforming the baseline model by a large margin of 25% in accuracy, offering a practical solution for large-scale ontology mapping.

## 1   Introduction

Categorizing concepts and entities into a taxonomy is a fundamental task of digital libraries. A universal entity categorization system is desirable for various tasks such as entity disambiguation, entity linking, and text classification. For such a universal taxonomy, Wikipedia serves as an indispensable knowledge resource, because it is the largest online encyclopedia containing a wide range of articles on vast topics. Most of its entries (i.e., Wikipedia articles) can be considered as (semi-structured) representations of entities. For grouping related articles, Wikipedia provides three complementary mechanisms: Categories, lists and navigation templates, based on shared characteristics or topics of articles [18]. Categories are organized in a hierarchy and each Wikipedia article is assigned to at least one category. Lists provide a mean for manual categorization of articles and can include entities that do not have a Wikipedia page yet. Lists are more difficult to process automatically due to informal construction.

---
* Jiaxin Qin's current affiliation is United Automotive Electronic Systems Co., Ltd

Although lists and categories of Wikipedia provide rich resources for universal taxonomy, their vastness and open creation introduce wildness such as irregularities and redundancies, causing difficulties in automating categorization of new entities. DBpedia [21] is a project aiming at extracting structured information from Wikipedia and organizing into a large knowledge graph. The DBpedia ontology is the heart of DBpedia, currently enlisting 788 classes, and around 4.8 million instances, on which a subsumption hierarchy is formed.

An ontology is a structured representation of knowledge that models a collection of concepts within a particular domain, along with their interrelationships [8]. However, independent development of ontologies often results in heterogeneous knowledge representations with varying categorizations and naming schemes. Ontology alignment is a vital solution to the semantic heterogeneity problem, which aims to establish the relation between semantically related entities in heterogeneous ontologies.

CaLiGraph [5, 6, 7] is a large semantic knowledge graph with a rich ontology compiled from the DBpedia ontology and Wikipedia categories and list pages, as depicted in Fig. 1(a). CaLiGraph expands the DBpedia ontology classes with fine-grained value restrictions to more than one million classes and over 200,000 restrictions by Cat2Ax [7] approach, which extracts entities from Wikipedia listings through a combination of the ontological information, axioms, and transformer-based extractors.

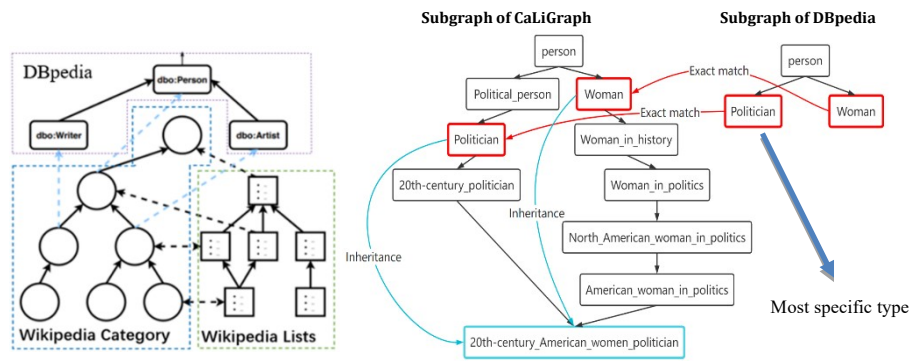

**Fig. 1 (a)** CaLiGraph ontology schema [5] **(b)** A mapping between CaLiGraph and DBPedia

Fig. 1(b) shows an example of mapping CaLiGraph classes to DBPedia classes. Suppose we intend to find a DBPedia class for the CaLiGraph class "20th-century_Americal_women_politician." The word "politician" indicates the possibil-ity of mapping the class to "Politician" in DBPeida, which can be done by lexical analysis. We also notice that the ancestors of the class also include "Politician," where class inheritance can be considered. However, the class "Woman" is also a candidate because of the word match, where multiple inheritance to the same class is occurring. In this case, "Politician" can be chosen as the most specific class because this is less populated than class "Woman."

Although the objective of CaLiGraph occupies a vital role in utilizing Wikipedia as a universal taxonomy, the quality of precise categorization on Wikipedia entities still needs improvement. In this paper, we propose an effective method for finding the most specific and accurate DBpedia class for a given Wikipedia list or category.

Our method includes the following parts: (1) A mapping method by lexical properties, which includes root phrase matching between class names of CaLiGraph and DBPedia, part-of-speech (POS) tagging, and inheritance in the class hierarchies. These mappings are then used for finetuning the pretrained language model (PLM) BERT [1]. Also, class name embeddings generated by SimCSE [4] are used to find semantically similar class names. We further utilize named entity typing on article titles of a target Wikipedia list or category, to find majority types in these listings. These approaches are combined to build CaLiGraph-DBpedia class pairs. (2) Finetuning of the pre-trained language model BERT, with the distantly supervised samples from the previous step. Here we also try to finetune BERT using a simple prompt-tuning strategy [10]. To reflect hierarchical properties of ontologies in prompt-based tuning, we use hierarchical classification such that ancestor class names are appended into prompt templates to enrich contextual information.

For evaluating the correctness of the generated mappings, we construct a manually annotated benchmark dataset, which consists of 3000 fine-grained CaLiGraph-DBpedia mapping pairs annotated by three annotators. We evaluate our model SLHCat on the CaLiGraph-DBpedia mapping task, by macro and micro F1-scores, and accuracy. Our proposed model outperforms the baseline model Cat2Ax by all the evaluation metrics by a large margin.

## 2 Related Work

### 2.1 Ontology alignment

Lexical matching serves as the foundation for traditional ontology mapping solutions, which is often combined with structural matching. This gave rise to various existing systems such as Cat2Ax [7]. Their lexical matching approach, however, only focuses on the text's surface form, such as overlapping sub-strings and sharing a textual pattern, which is unable to capture word semantics. Lexical and structural matchings have recently been suggested to be replaced by machine learning; for instance, DeepAlignment [9] and OntoEmma [16] use word embeddings to represent classes and compute the similarity of two classes according to the Euclidean distance between their word vectors. However, these approaches either require extensive feature engineering that is ad-hoc and relies on great amounts of annotated examples for training, or they use classic non-contextual word embedding models like Word2Vec [11], which only learns a global (context-free) embedding for each word.

### 2.2 CaLiGraph and Cat2Ax

CaLiGraph is a large semantic knowledge graph that incorporates a rich ontology compiled from the DBpedia ontology and Wikipedia categories and list pages [6]. The ontology is enriched with fine-grained value restrictions on its classes that are discovered with the Cat2Ax approach. CaLiGraph covers over 1 million classes and over 200,000 restrictions, containing information of 15 million entities, as shown in Table 1.

Table 1. Statistics of DBpedia and CaLiGraph

|  | Classes | Instances |
|---|---|---|
| DBpedia | 788 | 4,828,418 |
| CaLiGraph | 1,061,597 | 15,230,974 |

The following four major steps are used in Cat2Ax for allocating a DBPedia class to Wikipedia lists and categories: (1) Identify candidate category sets that share a textual pattern. (2) Find characteristic properties and types for candidate sets and combine them to patterns. (3) Apply patterns to all categories to extract axioms. (4) Apply axioms to their respective categories to extract assertions. The estimated mapping accuracy of Cat2AX reported in [7] is 96.8%. However, the authors' independent examinations on the mappings generated by Cat2Ax, reported in Section 5.1, reveals that 50% of mappings are either onto very general DBpedia types, mapping to a wrong type, or no DBpedia type is given. One example is that the CaLiGraph class "Male actor from Saskatchewan" is mapped to DBpedia class "Person" by Cat2Ax, but in this case "Actor" is more specific and appropriate.

### 2.3 Distant Supervision

Most of machine learning and deep learning techniques require a large amount of training samples for training a model for the target task. Manual labeling of training data requires considerable time and cost. An alternative approach to annotating training data is distant supervision [12], in which training samples are labeled automatically based on certain rules, suitable for situations where training data construction is costly.

Distant supervision for semantic typing is an extension of the paradigm used by [15] for utilizing WordNet to uncover hypernym (is-a) relations between entities, and is analogous to the application of poorly labeled data in bioinformatics [1][13], and in relation extraction which has no labeled data [12]. A typical distant supervision's assumption for relation typing is that any statement that contains two entities that are involved in a relationship may refer to that relationship [12].

In our work, we introduce distant supervision rules for discovering mappings between CaLiGraph classes and DBpedia classes, where the rules are based on textual and semantic similarities between class names, knowledge graph structures, and the lexical database WordNet on semantic relations.

### 2.4 Prompt-based learning

Prompt-based learning [15] is a new paradigm for finetuning PLMs for specific tasks by providing task-specific prompts, which has attracted attentions in the NLP field. In prompt learning, downstream tasks are formalized as equivalent cloze-style tasks by adding some pieces of text as a prompt, and the PLM is asked to handle these cloze-style tasks instead of original tasks. In contrast to the traditional finetuning methods, prompt-based learning does not require extra neural layers and bridges the objective form gap between pre-training and finetuning.

## 3 Problem definition and overview

In this paper, we focus on finding the most specific and matching DBpedia class for each of Wikipedia category and list names included in CaLiGraph. This task can be formulated as a multi-class classification problem.

Ontologies are often composed of statements in the form of triples (subject, predicate, object) in the Web Ontology Language OWL. Here, we denote by $O$ and $O'$ the two ontologies of CaLiGraph and DBpedia, respectively. The named class set $O$ (resp. $O'$) is denoted as $C$ (resp. $C'$). The set of articles assigned to a class $c \in C$ is denoted as as $e_c \in E_c$.

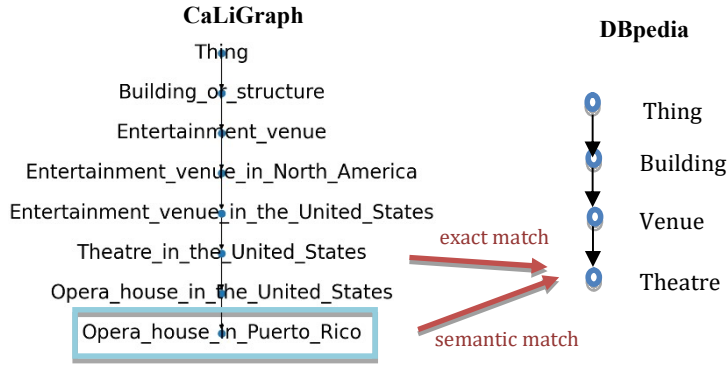

**Fig. 2.** Matching by exact and semantic match on root phrases

The goal of this paper is to find a mapping that maps each class $c \in C$ to a target class $c' \in C'$, where the target class $c'$ should represent the minimum concept that subsumes $c$. In other words, our goal is to find the most specific named class $c'$ of DBpedia for each CaLiGraph class $c$.

Fig. 2 shows an example of the task of mapping CaLiGraph classes to DBPedia classes. Suppose that "Opera_house_in_Puerto_Rico" is the target CaLiGraph class. Lexical analysis of the class name indicates the *root phrase*, which is the phrase located at the root of its dependency tree. The phrases "house," "opera house," "house in Puerto rico" are also root phrases. Using this root phrase set, we can search the DBPedia classes. However, since there is no exactly matching class in DBpedia, we need to consider semantic similarities between class names. DBPedia classes "Venue" and "Theatre" are semantically close to "opera house." For evaluating semantic similarities, we can utilize contextualized representations of texts generated by PLMs, such as SimCSE [4] and BERT [2]. "Theartre" turns out to be most specific and semantically close to the target, so "Theartre" will be selected.

An alternative approach to the above is utilizing inheritance on the class hierarchies. The target class has an ancestor "Theatre_in_the_United_States." The root phrase of this class name is also matching with "Theatre." We can propagate this class mapping to its descendants. The approach of hierarchical classification [15] can be applied here. One another approach is utilizing named entity typing. Existing part-of- speech (POS) tagging tools can provide named entity typing. Although the results of the typing are general, such as "house" is typed as "building," this extra information can assist semantic matching by PLMs.

## 4 Methodology

### 4.1 Model architecture

In this paper, we propose a novel ontology mapping method **SLHCat**, which exploits the hierarchical structures of DBpedia and CaLiGraph, as well as lexical and semantic similarities between class names. Then a hierarchical classifier based on pre-trained language model BERT is finetuned by distantly-supervised samples. Fig. 3 shows the overall structure of SLHCat. Our method consists of two main parts: (1) Generating distantly-supervised training samples. (2) Finetuning pretrained language models.

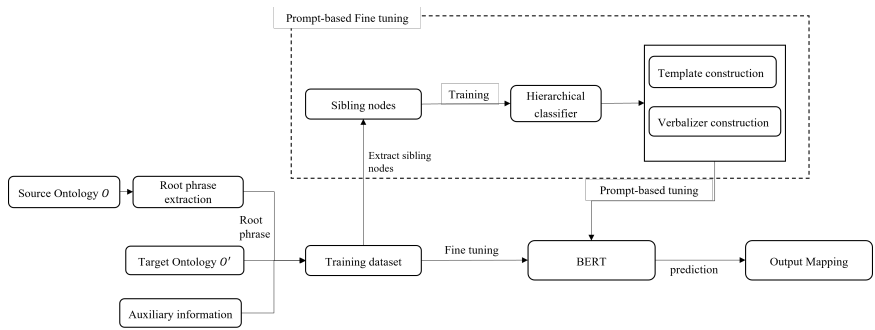

**Fig. 3.** Overall structure of our proposed model SLHCat.

### 4.2 Root phrase extraction

A DBPedia class node in the hierarchy of DBPedia may have an exactly identical class name in the hierarchy of CaLiGraph. As shown in Fig. 1(a), CaLiGraph uses DBPedia as an upper-level taxonomy and categorizes these rather general types in DBPedia into more specific types, although the mapping is not accurate enough. We check an exactly matching CaLiGraph node in the hierarchy for every DBPedia node. For example, "Joan Baez compilation album" is a CaLiGraph node, which contains "album" as a part of the class name, which is also found in DBpedia. So, it is likely that "Joan Baez compilation album" is mapped to "Album". Such matching pairs can be utilized as confident samples for distant supervision.

More generally, we define the *root word* as a single noun word which represents the subject information of the whole sentence. For a long noun phrase, the root word holds the most basic meaning of the whole phrase. However, considering only the root word may not be sufficient to find matching class names. The root word needs to be extended to a phrase by adding part of words that are modifying the root word. A *root phrase* is a phrase consisting of a root word and its modifying components. Fig. 4(a) shows an example of part-of-speech (POS) tagging and Fig. 4(b) shows an example of dependency parsing for class name "American football team in Finland". Here, the phrase "football team" is more specific than the root word "team," which could be matched with DBPedia class "SportTeam." But the most specific DBPedia type for this CaLiGraph class is "AmericanFootballTeam." Also, the phrase "team in Finland" can be a candidate of matching class names. This example indicates that we need to extract a candidate set of root phrases from a given class name, and then select the root phrase that matches with the most specific and semantically consistent and DBpedia class.

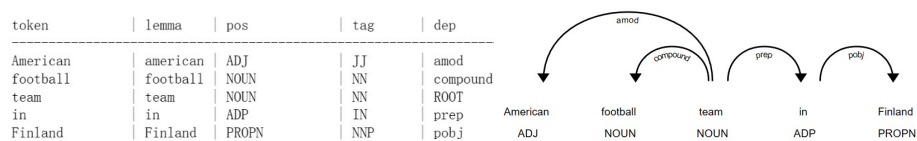

**Fig. 4.** (a) "American football team in Finland" after POS tagging  (b) dependency parsing

We extract a root phrase candidate set for each CaLiGraph class as follows:
**Step 1.** Perform POS tagging and dependency parsing on the CaLiGraph class names. Extract root words and append them to the root phrase set.
**Step 2.** In the dependency parse tree, enumerate the left and right subtrees of the root word, and add the phrase corresponding to each subtree to the root phrase set.
**Step 3.** If the subtree node is a preposition, append the preposition and its descendant nodes to the root word, and add the corresponding phrase to the root phrase set.

In the example of "American football team in Finland," we obtain the root phrase candidate set consisting of "team", "American team", "football team", "American football team" and "team in Finland."

SpaCy [20] is a tool for advanced natural language processing, which provides modules for POS tagging and dependency parsing. In this paper, we utilize these two modules to analyze the dependency structure of sentences and extract root words and root phrases.

### 4.3 Sentence embedding for matching by semantic similarities

Sentence embedding can capture semantic and contextual features of texts and the embedding can be mapped onto a shared vector space, on which semantic similarities can be measured by certain distance metrics over the vector space. In this paper, we use unsupervised SimCSE [4] to embed each root phrase $v_{root\_phrase}$ in the root phrase candidate set $V_{root\_phrase}$, where $v_{root\_phrase} \in V_{root\_phrase}$, and DBpedia class names $V_{DBO}$. We utilize the cosine similarity to calculate the distance between each root phrase from the root phrase candidate set and DBpedia class names as follows:

$$sim(v_{root\_phrase}, V_{DBO}) = \cos(v_{root\_phrase}, V_{DBO}) = \frac{\overrightarrow{v_{root\_phrase}} \times \overrightarrow{V_{DBO}}}{|v_{root\_phrase}| \times |V_{DBO}|}$$

We choose the pair of a root phrase and DBpedia class which has the highest cosine value as a confident sample for BERT finetuning under distant supervision. In the previous example, the cosine value of the root phrase "American football team" and DBpedia class "AmericanFootballTeam" achieves the highest, which is close to 1, so we assign "AmericanFootballTeam" to CaLiGraph class "American football team in Finland."

There may exist multiple pairs whose cosine values are close to 1. In such cases, we choose the DBpedia class which has the longest word matching to the CaLiGraph class name. For the CaLiGraph class "Collectible card game", the cosine values of root phrases "game" and DBpedia class "Game", "card game" and "CardGame" are close to 1. In this situation, we choose "CardGame" as the corresponding DBpedia class of "Collectible card game".

### 4.4 Propagating classes through CaLiGraph hierarchy

The subsumption hierarchies of CaLiGraph and DBpedia can be utilized for class inference. Suppose that a confident mapping that maps a CaLiGraph class $c \in C$ to a DBPedia class $c' \in C'$. Then we can extend the mapping to the descendants of $c$ as type inheritance. Regarding the example of Fig. 2, CaLiGraph class "Theatre_in_the_United_States" can be mapped "Theatre", thorough matching on the root phrase. Then by class inheritance, "Opera_house_in_Puerto_Rico" can be also mapped to "Theatre." Class inheritance is useful when no corresponding DBpedia class is found for a descendant of    However, the hierarchies of Wikipedia lists and categories allow one class having multiple parents, causing multiple inheritance, as we saw in Fig. 1(b). To resolve multiple inheritance and choose one target class, we utilize the BERT classifier.

We call nodes sharing the same parent as sibling nodes. We observe that sibling nodes sharing the same root phrase are likely to have the same DBpedia type, thus propagating types to such sibling nodes is possible. Sibling nodes can be searched from each CaLiGraph node in the current confident labeled dataset. In Fig. 5, the classes "Penn State Lady Lions basketball player", "Virginia Tech Hokies women's basketball player" and "Virginia Tech Hokies women's basketball player" are the children of "College women's basketball player in the United States," and share the same root phrase "basketball player." These class names can be added to the confident dataset. This can be viewed as a form of class inheritance, but sharing a root phrase between the parent and siblings is a strong indicator that they belong to the same class, giving a higher priority when resolving multiply assigned classes.

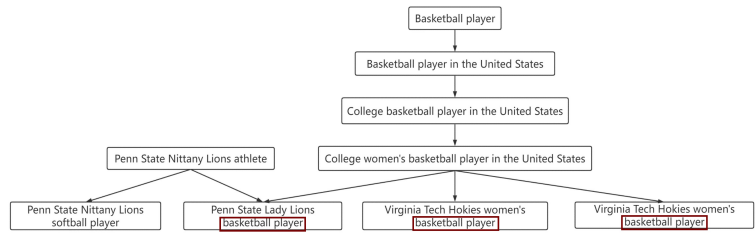

**Fig. 5.** Propagating classes to sibling nodes sharing a common root phrase

### 4.5 Named entity typing on Wikipedia article titles

Wikipedia articles included in the same category or list are supposed to share the same theme or subject. Thus, we can assume that in a given Wikipedia category or list, its member articles share a common aspect or attribute, from which we can infer the DBpedia class of the list or category.

**Fig. 6.** Wikipedia pages in category "American football teams in Finland"

Fig. 6 shows the Wikipedia articles belong to CaLiGraph class "American football team in Finland," on which several entities can be linked to American football players. To infer the DBpedia type of each member article, we utilize the technique of named entity typing on the article titles, where each article is regarded as representing a Wikipedia entity. Here we use the named entity recognition tool of SpaCy to identify the types of articles. The type of "Helsinki Roosters" is "ORG", which refers to companies, agencies, institutions, etc. We adopt a criterion such that we select the type that appears for more than half of the occurrences to reduce the potential errors in named entity recognition or the presence of multiple entity types within a single category or list.

Since the named entity types predicted by SpaCy do not exactly match with DBpedia classes, we map the types to DBpedia classes based on their descriptions. Table 2 shows the named entity types, their descriptions, and the mapped DBpedia classes.

Table 2. The mapping from named entity types to DBpedia class

| Named entity type | Description | DBpedia class |
| --- | --- | --- |
| PERSON | People, including fictional. | Person |
| NORP | Nationalities or religious or political groups. | Organization |
| ORG | Companies, agencies, institutions, etc. | |
| FAC | Buildings, airports, highways, bridges, etc. | ArchitecturalStructure |
| GPE | Countries, cities, states. | |
| LOC | Non-GPE locations, mountain ranges, bodies of water. | Place |
| PRODUCT | Objects, vehicles, foods, etc. (Not services.) | Thing |
| EVENT | Named hurricanes, battles, wars, sports events, etc. | Event |
| WORK_OF_ART | Titles of books, songs, etc. | Work |

This method can be effective when member article titles indicate entity types such as person names, which can complement the approaches based on ontology hierarchies and class names. However, the resulting DBpedia classes shown in Table 2 are rather general, locating around the top level of DBpedia. To find more specific and precise DBpedia classes, we need to search descendants of these classes.

Similarly to named entity typing on classes names, we can utilize the external lexical database WordNet for typing root words. WordNet provides semantic relations between words and sets of cognitive synonyms (synsets), as well as POS tags having suffix (POS.suffix), such as *noun.person* and *noun.animal*, from which we can extract candidate DBpedia classes. For example, the root word of CaLiGraph class "Recipient of French pardons" is "recipient", and the corresponding POS.suffix is noun.person. The POS tags and lexical name can further give us hints of labeling. However, as the POS tags from WordNet are either too general, or not fitting well with DBpedia classes, in this work we only consider the types *noun.person* and *noun.group*. If the root word of a CaLiGraph class name belongs to *noun.person*, it is highly likely that the CaLiGraph class belongs to DBpedia class "Person" or its descendants.

### 4.6 Resolving results predicted by multiple methods

In the step of Section 4.3 for obtaining vector representations using SimCSE, there can exist multiple vector pairs with similarity close to 1, which means that CaLiGraph classes have a nearly exact match to certain DBpedia classes. We regard such CaLiGraph-DBpedia pairs as having high confidence, and use these pairs for distant supervision. For those whose similarity is significantly less than 1, we use the candidate types obtained in the previous steps, and select a DBpedia class according to the following rules:

Rule 1: If the WordNet POS.suffix of the root word is *noun.person* or *noun.group* we only consider the types under person or group, respectively.

Rule 2: Suppose the DBpedia classes predicted by 1) named entity typing, 2) CaLiGraph hierarchy, and 3) sentence similarity with similarity score higher than 0.75 form a directed path in the DBpedia hierarchy. Then we choose the DBpedia class that is deepest in the path.

Rule 3: If two or more methods predict one identical class, the class is selected.

Rule 4: If none of the above rules are met, the class given by named entity typing is selected.

### 4.7 BERT Finetuning

**BERT finetuning.** Given sets of CaLiGraph-DBpedia pairs generating from previous steps, we finetune a pre-trained BERT model along with a downstream binary classifier on the cross-entropy loss. We limit the inputs length of BERT to 256. The classifier consists of a linear layer which takes as input the embedding of [CLS] token from BERT's last-layer outputs and apply to the output softmax layer. The optimization is done by Adam algorithm. The final output is the probability distribution over the DBpedia classes for the given CaLiGraph class.

**Prompt-based tuning.** In our prompt-based learning, for each DBpedia class $c' \in C'$, we define a supplemental word set $S_{c'} = \{\omega_1, \omega_2, ..., \omega_m\}$, where $S_{c'}$ is a subset of the vocabulary of BERT. A prompt template $T(\cdot)$ wraps the input $c \in C$ (the input is CaLiGraph class) into a prompt input $T(c)$ by adding additional tokens, and a [MASK] token is added. The classification task is transformed into a masked language modeling (MLM) problem, that is to predict the missing word in [MASK] [3].

*Verbalizer.* The verbalizer is a significant part of prompt-tuned classification task, which projects the words that predicted at [MASK] to our label set. We use related words [19] to expand the DBpedia classes. Specifically, we chose the top-10 related words for each class in DBpedia as the supplemental word set of the class.

For MLM, we use the confidence scores of all the words in $S_{c'} = \{\omega_1, \omega_2, ..., \omega_m\}$ to construct the final score of each DBpedia class.

$$P(c'|c) = \frac{1}{m}\sum_{j}^{m} \lambda_j P([MASK] = \omega | T(c))$$

where $m$ is the number of top-ranked related words, $\omega$ is a related word for DBpedia class $c'$, and $\lambda_j$ is a parameter indicating the importance of the current word $\omega_j$.

*Template.* Template wraps the inputs with textual or soft-encoding sequence. Here, we utilize the soft-encoding templates, which can continuously optimize in a vector space. Soft-encoding strategy introducing special tokens $[P_1], ..., [P_l]$ as the templates. The template becomes:

$$T_1(c) = [P_1] \ldots [P_n] \, Category \, [P_{n+1}] \ldots [P_l].c.$$

**Hierarchical classification as hint for prompt-based learning.** Hierarchical classification [14] is an effective approach for classification when the target classes are organized in a tree structure. As we discussed in Section 1, we can proceed hierarchical classification according to the hierarchy of the DBpedia ontology, where the classification result at a class is utilized as a hint $H$ or bias at its child classes. The template that utilizes the result of hierarchical classification is realized by:

$$T_2(c) = [P_1] \ldots [P_n] \, Category \, [P_{n+1}] \ldots [P_l].c.H.$$

## 5 Experiments

### 5.1 Benchmark dataset construction

To evaluate the accuracy of the mappings of the proposed method, the authors constructed a benchmark dataset that consists of 3,000 mappings from CaLiGraph classes to DBpedia classes. The benchmark dataset construction was done as follows: 3,000 classes were randomly sampled from CaLiGraph. For each CaLiGraph class, one corresponding DBpedia class was selected by three annotators, where the most specific and valid class was instructed to be chosen. The mappings were cross-checked by the three annotators and one DBpedia class was selected after reaching agreement.

We compare the mappings generated by Cat2Ax and our benchmark dataset. Table 3 shows the results of comparison. We find that 49.2 % of the mappings generated by Cat2Ax are identical to our manual mappings, which we judge as correct. On the other hand, the remaining 17.8% are judged as wrong, in which 437 mappings are not specific enough. The remaining 32.9% classes have assigned no DBpedia class, meaning that Cat2Ax failed to find a class. Overall, 50% of CaLiGraph classes have not been assigned appropriate DBpedia classes, leaving room for improvement.

**Table 3** Evaluation of Cat2Ax mappings by the benchmark dataset

| Cat2Ax mapping | | Benchmark dataset | Percentage |
|---|---|---|---|
| Correct | | 1476 | 49.2% |
| Wrong | Not specific | 437 | 17.8% |
| | Others | 97 | |
| Missing | | 986 | 32.9% |
| Total | | 3000 | 100% |

For finetuing the PLM, we generated another 3,000 mappings by the distant supervision method described in Section 4, as follows: 3,000 CaLiGraph classes were randomly sampled, where the classes are disjoint from the 3000 classes used for the benchmark dataset. Then the method of Section 4 was used to assign DBpedia classes to construct 3,000 mappings. The training dataset is divided into 90% for the training set and 10% for the validation set. Then the training dataset was augmented into 12,700 classes, by propagation on sibling nodes described in Section 4.4. Table 4 shows the statistics of the benchmark dataset.

Table 4. Statistics of datasets

| Dataset | Training data | | Augmented set | Benchmark dataset |
|---|---|---|---|---|
| | Training set | Validation set | | |
| CaLiGraph -DBpedia | 2700 | 300 | 12700 | 3000 |

### 5.2 Experimental Settings

Our proposed model SLHCat, for mapping CaLiGraph classes to DBpedia classes, is evaluated over the benchmark dataset. SLHCat is compared against the baseline model Cat2Ax. We evaluate effectiveness of each proposed component, by comparing seven different configurations (a) – (g), where each configuration is shown in Table 5. Each symbol means that: (**Dist**) No PLM classifier, and only distant supervision rules are used. (**BERT**) BERT finetuning. (**Prompt**) Prompt-based tuning. The distant supervision rules are divided as: (**Root Phase**) Root phrase set. (**NER**) Name entity typing on entities in Wikipedia member pages. (**Inherit**) Class inheritance. (**Lex**) Lexical typing by WordNet. (**Hier**) Hierarchical classification approach in prompt tuning.

The evaluation metrics of the experiments are {Macro, Micro}-{Precision, Recall, F1 score} and Accuracy. For BERT finetuning, we use vanilla finetuning. We choose early stopping to monitor the training process and determine the early training stop when the performance on the validation set starts to deteriorate. In the training, when the validation loss no longer decreases for 7 consecutive times, the model is considered to have converged, and the training is stopped. The training batch size is equal to 16, the input max length is 256 and learning rate is 1e-5 for both vanilla finetuning and prompt-based tuning.

### 5.3 Results and discussions

Table 5 shows the results of the evaluation. The overall results demonstrate that our proposed method SCHCat achieved higher scores in all the seven metrics than the baseline model Cat2Ax. Within the seven configurations, using BERT finetuning for final classification is prevailing over the models (a) using only the distant supervision rules, and (f, g) prompt-based tuning. Full-finetuning of BERT achieves higher performance than the prompt-based approach, which is explained as prompt-based tuning is training only prompt-related parameters, although it is cost effective.

**NER** (c) is improving Micro and Macro F1 scores by 0.015 and 0.081, respectively, over the model (b) using only **Root Phrase**. **Inherit** (d) shows improvement over (c) on Micro-F1 score and accuracy, while **Lex** (e) shows improvement over (d) on Macro-F1 score. Macro-F1 score assigns equal weights on both populated and minority classes in averaging, indicating that **Lex** is showing effectiveness on minority classes.

Same trends are observed between **Hier** (g) and without **Hier** (f) on the prompt-based models. In terms of accuracy, the prominent configuration of SCHCat is (d) **BERT**+**Root Phrase** + **NER** + **Inherit** + **Lex**, improving the accuracy of Cat2Ax by 0.251.

**Table 5** Results of mapping categories and lists to DBpedia classes by SLHCat and baseline

| Model | Macro-Pre | Macro-Recall | Macro-F1 | Micro-Pre | Micro-Recall | Micro-F1 | Accuracy |
|---|---|---|---|---|---|---|---|
| **Cat2Ax** | 0.342 | 0.335 | 0.321 | 0.528 | 0.492 | 0.509 | 0.492 |
| **SLHCat** | | | | | | | |
| a) Dist + Root Phrase + NER + Inherit + Lex | 0.623 | 0.567 | 0.562 | 0.670 | 0.662 | 0.666 | 0.662 |
| b) BERT + Root Phrase | 0.607 | **0.667** | 0.598 | 0.699 | 0.642 | 0.669 | 0.642 |
| c) BERT + Root Phrase+ NER | **0.646** | 0.647 | 0.613 | 0.746 | 0.723 | 0.734 | 0.723 |
| d) BERT + Root Phrase + NER + Inherit | 0.625 | 0.658 | 0.608 | **0.774** | **0.743** | **0.758** | **0.743** |
| e) BERT+ Root Phrase + NER + Inherit +Lex | 0.636 | 0.663 | **0.618** | 0.753 | 0.723 | 0.728 | 0.723 |
| f) Prompt + Root Phrase + NER + Inherit + Lex | 0.601 | 0.623 | 0.578 | 0.748 | 0.712 | 0.729 | 0.712 |
| g) Prompt + Root Phrase + NER + Inherit + Lex + Hier | 0.644 | 0.627 | 0.607 | 0.741 | 0.703 | 0.721 | 0.703 |

## 6  Conclusion and Future work

In this paper, we proposed a novel approach for ontology alignment that utilizes distant supervision to automatically generate confident mappings, for finetuning a pre-trained language model. Our approach covers textual, semantic, lexical, and structural features of ontologies. We employ two different training strategies, namely prompt-based tuning and finetuning on the pre-trained language model BERT. Hierarchical classification is employed to give guiding signals in prompt-based tuning. To evaluate the correctness of mappings, we constructed a benchmark dataset consisting of 3,000 labeled mappings, through manual annotation. Our proposed method outperforms the baseline Cat2Ax by a wide margin of 0.25 in accuracy.

While category and list structures of Wikipedia are considered as an important source for finding valid mappings, considerable noises are also introduced due to not well-maintained link structures. In future work, we shall consider denoising mechanisms to improve correctness, and consider utilization of large language models, for augmenting class names.